\documentclass[conference]{IEEEtran}
\IEEEoverridecommandlockouts

\usepackage{cite}
\usepackage{amsmath,amssymb,amsfonts}
\usepackage{algorithmic}
\usepackage{algorithm}
\usepackage{graphicx}
\usepackage{subfig}
\usepackage{subcaption}
\usepackage{textcomp}
\usepackage{xcolor}
\usepackage{booktabs}
\usepackage{multirow}
\usepackage{soul}
\def\BibTeX{{\rm B\kern-.05em{\sc i\kern-.025em b}\kern-.08em
    T\kern-.1667em\lower.7ex\hbox{E}\kern-.125emX}}
\begin{document}

\title{SparsePilot: Belief-Guided Network Planning under Sparse Wireless Measurements\\
}

\author{
    \IEEEauthorblockN{
        Xuanhao Luo\IEEEauthorrefmark{1},
        Jiayuan Huang\IEEEauthorrefmark{1},
        Longyu Zhou\IEEEauthorrefmark{2},
        Mingzhe Chen\IEEEauthorrefmark{3},
        Yuchen Liu\IEEEauthorrefmark{1}
    }
    \IEEEauthorblockA{
        \IEEEauthorrefmark{1}North Carolina State University,
        \IEEEauthorrefmark{2}Singapore University of Technology and Design,
        \IEEEauthorrefmark{3}University of Miami
    }
}

\maketitle
\begin{abstract}
Unmanned aerial vehicles (UAVs) have emerged as a promising solution for on-demand wireless coverage planning in urban environments. Existing learning-based UAV control methods, however, typically rely on continuous access to dense user-level received signal strength (RSS) measurements. Such full-observation assumptions are difficult to satisfy in real-world deployments due to the high cost and limited availability of dense wireless feedback. Sparse-feedback decision making under severe observation constraints therefore represents a fundamental challenge. To fill this gap, we propose SparsePilot, a measurement-efficient sensing-control framework that couples active wireless probing with belief-guided network control. SparsePilot formulates spatial probing as a multi-armed bandit problem over grid cells, uses upper confidence bound probing to select informative regions, and aggregates sparse RSS measurements into a coverage belief map. A deep reinforcement learning controller then uses this belief state to generate continuous UAV mobility actions, while the full wireless state remains hidden from the policy.
We further provide a theoretical analysis connecting sparse probing, belief estimation error, and the sparse-feedback performance gap. Experiments across seven urban digital twins show that SparsePilot achieves superior coverage restoration performance while using only about 3.1\% of the full-observation measurement budget and demonstrates strong cross-scene generalization to unseen urban-scale wireless environments.
\end{abstract}


\section{Introduction}
Unmanned aerial vehicles (UAVs) have emerged as flexible wireless infrastructure for coverage extension, hotspot support, emergency communication, and rapid network recovery \cite{mozaffari2019tutorial, li2018uav}. Compared with fixed terrestrial base stations, UAVs can be dynamically repositioned to compensate for weak-coverage regions caused by urban blockage, uneven user distribution, temporary traffic demand, or sudden infrastructure degradation \cite{alzenad20173}. This flexibility is particularly valuable in complex urban environments, where propagation conditions vary significantly across space and UAV placement often requires iterative adjustment based on updated wireless feedback, such as channel state measurements and link quality indicators~\cite{nguyen20223d}.

Wireless digital twins (DTs) provide an effective platform for designing and evaluating UAV-assisted networks before real-world deployment \cite{testolina2024boston, lin20236g}. By combining city-scale three-dimensional (3D) scene models, base-station configurations, user distributions, and ray-tracing-based propagation modeling, a DT can estimate received signal strength (RSS), coverage quality, and service reliability under different network configurations \cite{he2023physics}. Such a platform enables closed-loop optimization of UAV positions without disrupting the physical network. As a result, recent studies have increasingly explored deep reinforcement learning (DRL) for UAV placement, trajectory optimization, and coverage control in simulated or DT environments \cite{hoang2025adaptive, zhan2025joint}.

Despite this progress, many learning-based UAV control formulations rely on a full-observation assumption \cite{hoang2025adaptive}. Specifically, the controller is often assumed to observe complete ground-user locations, RSS values, or channel states at \textit{every} decision step. Although such information is available in a simulator, a deployed wireless DT cannot be assumed to remain perfectly synchronized with the physical network. Maintaining an up-to-date DT requires continuous data exchange, and communication delay, packet loss, and limited wireless resources can create synchronization latency and state mismatch \cite{cakir2023synchronize, li2025delay}. In practice, these wireless measurements must be collected through probing signals, user reports, or base-station feedback. Such measurements introduce signaling overhead, measurement latency, and device-side energy cost \cite{li2025contextual, giordani2018tutorial}. Moreover, idle or disconnected users may be unable to continuously report their channel conditions. As a result, the full-observation policies often overestimate the information available to a deployable network controller.

This practical issue motivates a fundamental question: \textit{how can network control systems make effective decisions when the global wireless state is not directly observable?} In this paper, we study sparse-feedback UAV-assisted network planning, where the controller observes only a small subset of user- or region-level RSS measurements at each decision epoch. 
Unlike conventional UAV placement or trajectory optimization, which assume complete network visibility, this setting inherently \textit{couples} sparse wireless sensing with network control in the same loop: the controller must actively acquire informative measurements from an incomplete belief of the wireless environment while simultaneously making UAV deployment decisions under partial observability.

To this end, we propose SparsePilot, a measurement-efficient sensing-control framework for sparse-feedback network planning. SparsePilot divides the service region into grid cells and formulates sparse wireless probing as a multi-armed bandit (MAB) problem over spatial regions. Each grid cell is treated as a probing arm, whose feedback is used to estimate regional coverage deficiency and uncertainty. Then, an upper confidence bound (UCB) strategy actively selects informative cells that are either likely to be under-covered or have high sensing uncertainty. The collected RSS measurements are then aggregated into a three-channel coverage belief map, where each cell encodes its estimated outage ratio, RSS deficiency, and sensing uncertainty.
At the second stage, the constructed belief map serves as the observation for continuous UAV mobility control. We train a DRL-based policy with soft actor-critic (SAC) \cite{haarnoja2018soft} to map the belief-guided state to UAV displacement actions. In this design, UCB and SAC work synergistically: UCB decides \textit{where} to collect sparse measurements, while SAC decides \textit{how} to reposition UAVs based on the updated belief of the wireless environment. The full wireless state remains hidden from the policy and is used only for training reward computation and evaluation. This separation allows SparsePilot to retain closed-loop adaptation while operating under limited wireless feedback.
The main contributions of this paper are summarized as follows.
\begin{itemize}
\item We formulate a sparse-feedback 3D network planning problem using wireless belief maps, where the network controller cannot access the full RSS and coverage state and must operate under a limited probing budget.

\item We design SparsePilot, a measurement-efficient sensing-control framework that couples UCB-guided active probing, coverage belief construction, and SAC-based continuous UAV mobility control.

\item We theoretically analyze the impact of sparse probing on belief map estimation and belief-guided network control. The analysis characterizes how probing frequency affects cell-level belief error, how UCB allocates measurements to informative regions, and how belief error bounds the sparse-feedback mobility control gap.

\item We implement SparsePilot in ray-tracing-based wireless digital twins and evaluate it across multiple city-scale environments, showing strong coverage restoration performance while using only about 3.1\% of the full-observation measurement budget.

\end{itemize}

\begin{figure}[t]
	\centerline{\includegraphics[scale=0.3]{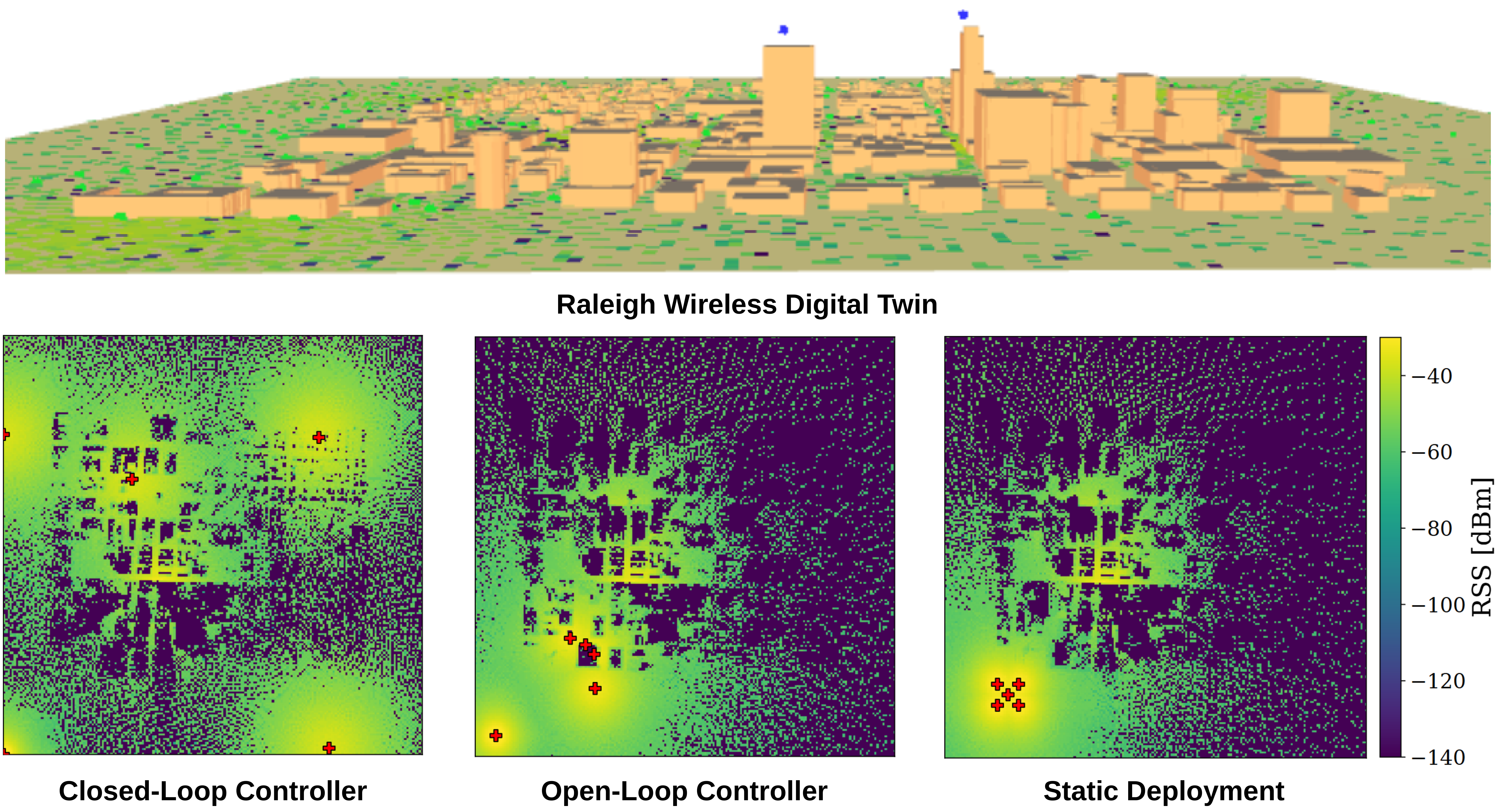}}
    \caption{\footnotesize RSS maps under different UAV deployment strategies. The DT can compute dense RSS maps, but a deployable controller cannot directly observe the full RSS/coverage state.}
    \label{fig:motivation_overview}
    \vspace{-1.8em}
\end{figure}

\section{Motivation and Observations}
\label{sec:motivation}

\noindent\textbf{Observation 1:} \textit{Closed-loop wireless feedback substantially improves UAV-assisted network planning.}

Wireless DTs and ray-tracing simulators, such as Sionna RT~\cite{sionna} and Wireless InSite~\cite{wirelessinsite}, provide a physically grounded platform for evaluating UAV-assisted networks under different deployments. As shown in Fig.~\ref{fig:motivation_overview}, different UAV placement strategies produce clearly different RSS distributions and coverage patterns within the same Raleigh scene. These spatial variations show that UAV-assisted coverage restoration is highly sensitive to deployment decisions, since each UAV movement changes transmitter locations, propagation paths, and user associations, thereby affecting the resulting RSS and coverage state.

\begin{figure}[t]
    \centering
    \includegraphics[width=0.45\linewidth]
    {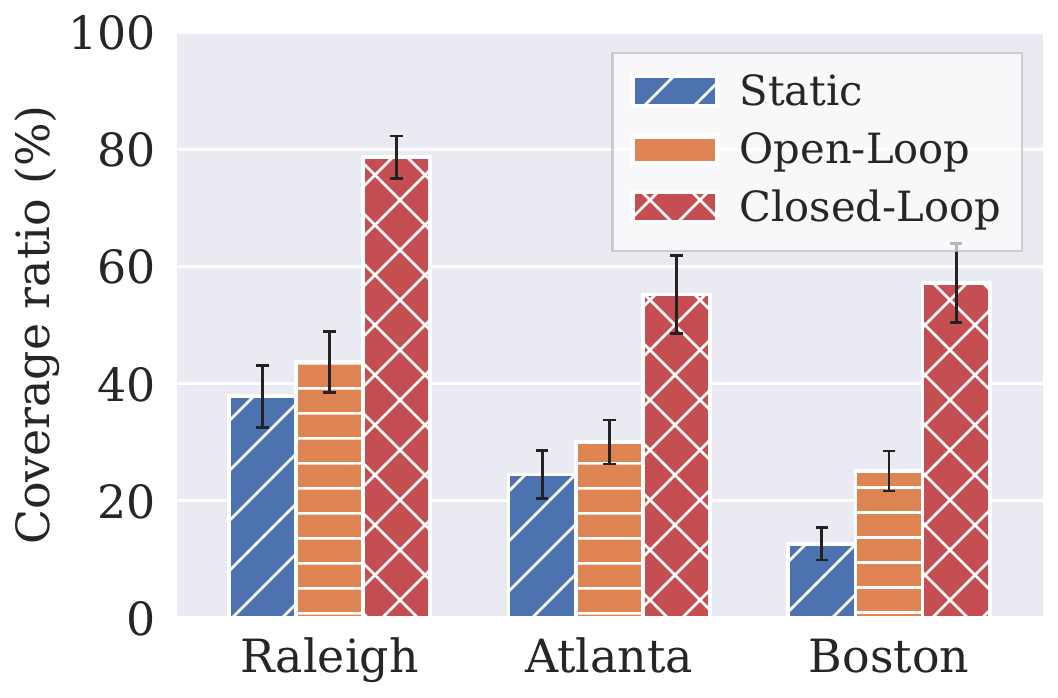}
    \hspace{-0.1em}
    \includegraphics[width=0.45\linewidth]
    {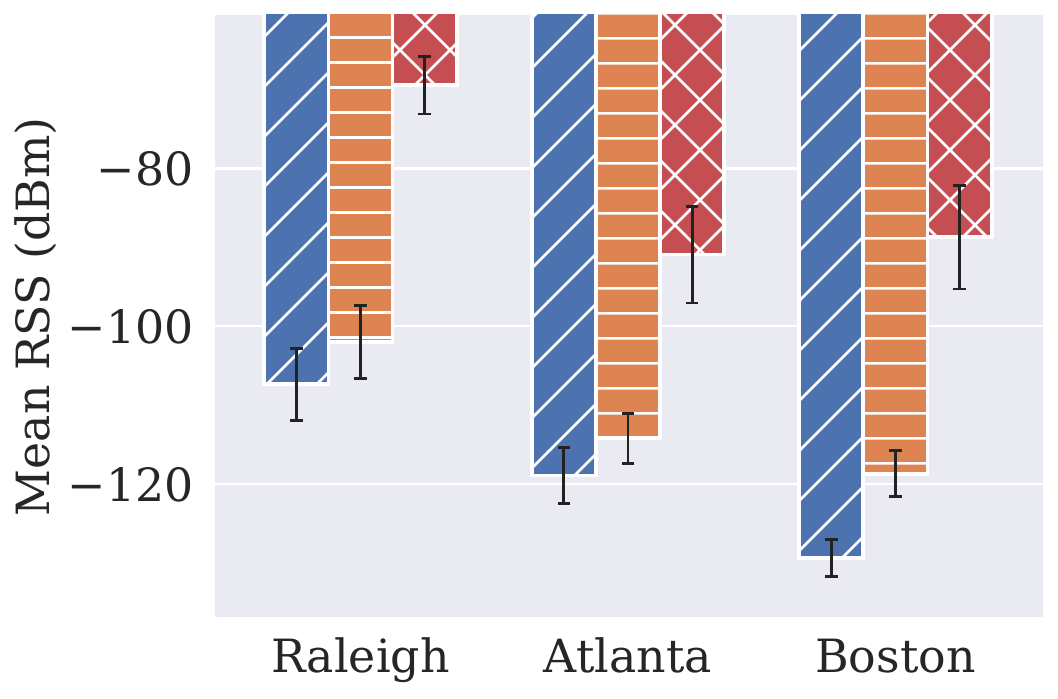}
    \caption{\footnotesize Performance comparison of different control policies. }
    \label{fig:motivation_gap}
    \vspace{-1.8em}
\end{figure}

To quantify whether online feedback improves such deployment decisions, we compare three representative settings in Fig.~\ref{fig:motivation_gap}. \emph{Static} keeps the UAVs at their initial locations. \emph{Open-loop control} makes a one-shot UAV movement decision without using online RSS or coverage feedback, similar to fixed-snapshot placement paradigms~\cite{lyu2016placement,qiu2020placement}. \emph{Closed-loop control} receives full user-level RSS and coverage feedback at each control step, following adaptive and sequential UAV control paradigms~\cite{hoang2025adaptive, nguyen20223d}. Across Raleigh, Atlanta, and Boston, open-loop control improves over the static policy, but remains consistently below dense-feedback closed-loop control in both coverage ratio and mean RSS. These results suggest that UAV-assisted coverage restoration is fundamentally a feedback-driven control problem, where online wireless feedback is essential for identifying coverage gaps and adapting UAV decisions to the evolving network state.

\noindent\textbf{Observation 2:} \textit{Complete user-level wireless feedback is not a deployable control interface.}

Although the closed-loop control enhances network planning, obtaining such feedback continuously in operational networks is challenging. 
Dense feedback requires collecting RSS or coverage reports from all users after each UAV movement, which introduces signaling overhead, reporting latency, device-side energy cost, and scheduling burden.
For example, with $N_u=150$ users and a planning horizon of $T=5$, dense feedback requires up to $150\times5=750$ user-step RSS measurements per episode. This burden scales linearly with the number of users and planning steps, making dense feedback costly in larger or frequently updated networks.

Therefore, neither open-loop nor full-observation control fully matches the operational constraints of urban-scale wireless networks. Open-loop approaches lack updated wireless feedback, whereas full-observation control assumes excessive measurement availability. A deployable controller must retain the benefits of closed-loop adaptation while probing only a small subset of users or spatial regions. Under this sparse-feedback setting, measurements must be allocated to informative regions, while UAV actions must be selected from incomplete observations. The central \textbf{challenge} is therefore to preserve the effectiveness of feedback-driven network planning under severe observation constraints.

\begin{figure*}[t]
	\centerline{\includegraphics[scale=0.689]{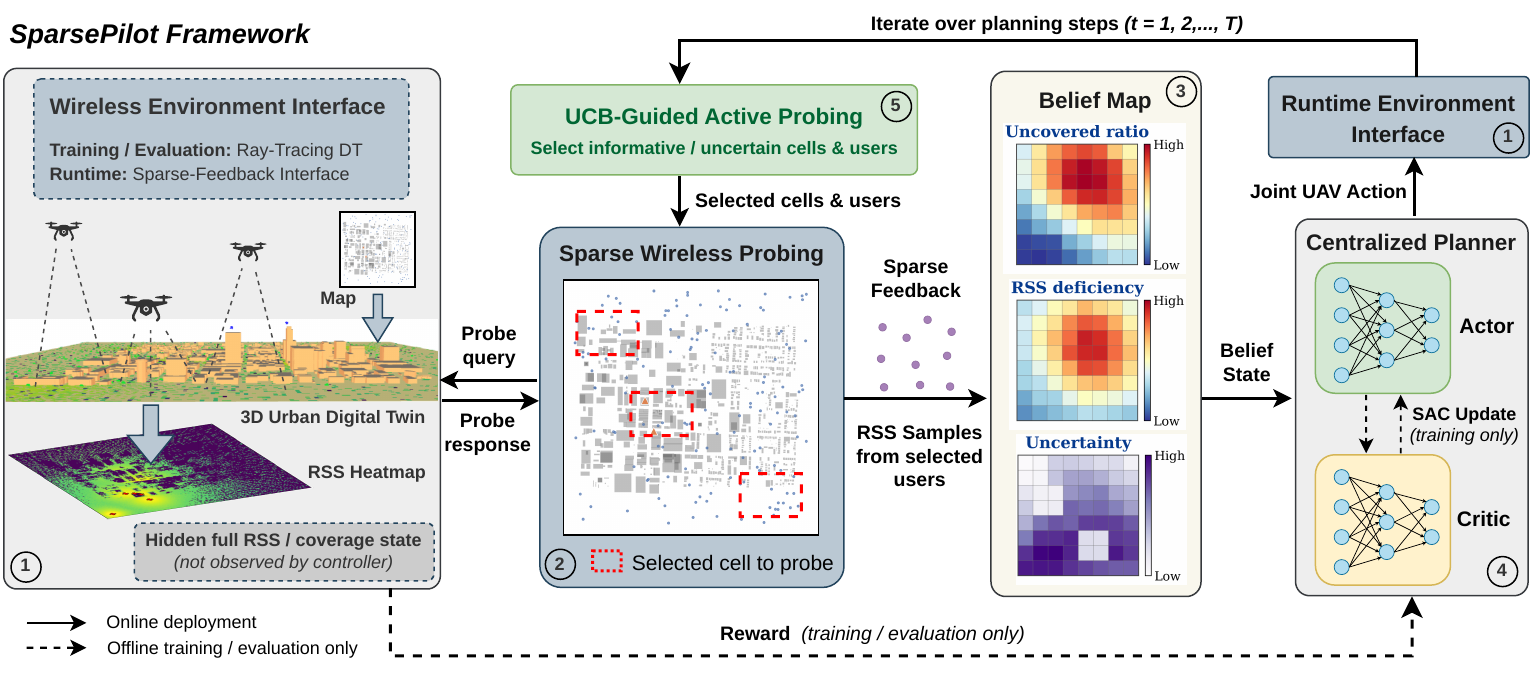}}
        \vspace{-0.5em}
    \caption{\footnotesize SparsePilot Framework Overview. The numbered labels indicate the closed-loop execution order: (1) the runtime environment interface provides sparse-feedback access, (2) sparse probing module collects selected wireless samples, (3) the belief map is updated, (4) the centralized planner outputs joint UAV actions, and (5) UCB-guided active probing continuously selects informative cells/user information for the next planning step. 
    } 
    \label{framework}
    \vspace{-0.5em}
\end{figure*}

\noindent\textbf{Implication:} \textit{Sparse-feedback UAV network planning requires the joint design of active sensing, belief construction, and network control.}

Rather than reconstructing a complete wireless state from limited measurements \cite{ren2026channel, luo2024rm, romero2022radio}, the network planner should maintain a task-oriented belief of the environment that preserves information most relevant to targeted coverage. Such a belief state enables the controller to reason about both estimated coverage deficiencies and measurement uncertainty, allowing UAVs to make informed decisions despite observing only a small fraction of the network. This shifts network planning from full-state optimization to belief-guided control under partial observability.

\section{SparsePilot System Design}
\label{sec:method}

To this end, we present SparsePilot, a joint sensing-control framework for sparse-feedback network planning under limited wireless measurements. As shown in Fig.~\ref{framework}, SparsePilot couples sparse wireless sensing and UAV mobility control through a compact coverage-based belief state. At each planning step, it selects informative spatial regions for probing, aggregates sparse RSS feedback into a grid-based belief map, and guides a belief-driven SAC controller to generate continuous actions.

\subsection{Sparse-Feedback UAV Planning Formulation}
\label{subsec:problem_formulation}

We consider a UAV-assisted coverage restoration problem in an urban wireless environment, while the underlying sensing-estimation-control framework can be extended to other network planning problems that require decision making from
sparse measurements and incomplete network visibility. The environment consists of a set of UAV-mounted transmitters $\mathcal{U}$, a set of fixed terrestrial base stations $\mathcal{B}$, a set of ground users $\mathcal{G}$, and a 3D urban scene with buildings and blockage. 

At each planning step, the UAVs can be repositioned to improve the downlink coverage of poorly served users. Let $r_{ij}(t)$ denote the RSS from transmitter $j \in \mathcal{U}\cup\mathcal{B}$ to user $i \in \mathcal{G}$ at step $t$. The best RSS of user $i$ is 
\begin{equation}
r_i(t) = \max_{j\in\mathcal{U}\cup\mathcal{B}} r_{ij}(t).
\end{equation}
Given a coverage threshold $\tau$, the binary coverage indicator of user $i$ is $z_i(t)=\mathbb{I}\{r_i(t)\geq \tau\}.$
The network coverage ratio is the fraction of covered users:
\begin{equation}
C_t = \frac{1}{|\mathcal{G}|}\sum_{i\in\mathcal{G}} z_i(t).
\end{equation}

Unlike a full-observation controller, SparsePilot probes only a small subset of spatial regions and users at each planning step. The remaining user-level RSS and coverage state is hidden from the controller. We assume that coarse user-location or cell-occupancy information is available for probe scheduling, for example through network-side localization, serving-cell records, or active user registration. This information is used only to identify populated cells and candidate users; the instantaneous RSS and coverage state of a user is observed only when that user is probed.
The objective is to learn a UAV mobility control policy that enhances network coverage under a limited wireless measurement budget. SparsePilot addresses this sparse-feedback control problem by integrating 1) UCB-guided probing, 2) grid-based coverage belief construction, and 3) belief-guided UAV control, as detailed below.

\subsection{UCB-Guided Sparse Wireless Probing}
\label{subsec:active_probing}

From an optimization perspective, sparse wireless probing cannot be directly solved as a conventional full-state optimization problem, since the channel conditions of different spatial regions are unknown before measurements are collected, and the effective coverage state evolves with UAV relocation.  We therefore formulate the probing process as an online bandit optimization problem and adopt UCB to efficiently allocate measurement budget between coverage-critical and uncertain regions through exploration and exploitation.

First, SparsePilot partitions the service region into an $N_g \times N_g$ grid, where each grid cell represents a spatial probing region. Each occupied grid cell is treated as an arm in an MAB problem. Pulling an arm corresponds to probing a small number of users inside that cell. At each planning step, the probing module selects at most $K_c$ grid cells and probes up to $K_u$ users in each selected cell. If the total number of users is $N_u$, the maximum per-step probing ratio is $\rho_{\max} = \frac{K_c K_u}{N_u}.$
This sparse probing budget is significantly smaller than full-observation, which requires measuring all users at every planning step.

For each grid cell $m$, SparsePilot maintains two statistics: the number of times the cell has been probed, denoted by $n_m(t)$, and an empirical coverage-deficiency score, denoted by $q_m(t)$. The coverage-deficiency score estimates how severely users in that cell suffer from outage or weak RSS. To balance exploitation of known weak regions and exploration of uncertain regions, SparsePilot assigns each occupied cell a UCB probing score:
\begin{equation}
\mathrm{UCB}_m(t)
=
q_m(t)
+
\beta
\sqrt{\frac{\log(t+2)}{n_m(t)+1}},
\label{ucb}
\end{equation}
where $\beta$ controls the exploration strength. The first term prioritizes cells that are likely to be under-covered, while the second term encourages probing cells with high uncertainty. At each step, SparsePilot selects the top-$K_c$ populated cells according to this score. Cells without active users are not considered probing candidates because they cannot return user-level RSS feedback in practice.

After selecting a cell $m$, the system probes a subset $\mathcal{P}_m(t)$ of users in that cell. Within each selected cell, if the number of candidate users exceeds $K_u$, SparsePilot uniformly samples $K_u$ users without replacement; otherwise, all users in the cell are probed. The selection does not use hidden RSS or coverage information. The measured uncovered-user ratio is
\begin{equation}
\bar{o}_m(t)
=
\frac{1}{|\mathcal{P}_m(t)|}
\sum_{i\in\mathcal{P}_m(t)}
\left(1-z_i(t)\right).
\end{equation}
The measured normalized RSS deficiency is
\begin{equation}
\bar{d}_m(t)
=
\frac{1}{|\mathcal{P}_m(t)|}
\sum_{i\in\mathcal{P}_m(t)}
\mathrm{clip}
\left(
\frac{\max(0,\tau-r_i(t))}{D_{\max}},
0,
1
\right),
\end{equation}
where $D_{\max}$ is a normalization constant. SparsePilot combines these two measurements into an instantaneous cell badness score:
\begin{equation}
\tilde{q}_m(t)
=
\alpha \bar{o}_m(t)
+
(1-\alpha)\bar{d}_m(t),
\label{badness}
\end{equation}
where $\alpha$ balances binary outage and RSS deficiency. 
The empirical cell score is updated by an incremental average:
\begin{equation}
    q_m(t+1)
    =
    \frac{
        n_m(t)q_m(t)+\tilde{q}_m(t)
    }{
        n_m(t)+1
    },
    \label{eq:q_update}
\end{equation}
and the probing count is updated as $n_m(t+1)=n_m(t)+1$. Here, the UCB module is only responsible for active measurement selection. 
It does not directly control UAV movement. UAV control is performed by a separate policy using the belief state constructed from these sparse measurements in Sec. \ref{subsec:belief_control}.

\subsection{Coverage Belief Map Construction}
\label{subsec:belief_map}
The sparse RSS measurements collected by the active probing module in Sec.~\ref{subsec:active_probing} are aggregated into a grid-based coverage belief map. For each grid cell $m$, SparsePilot maintains a three-channel belief vector:
\begin{equation}
b_m(t)
=
\left[
\hat{o}_m(t),
\hat{d}_m(t),
\hat{u}_m(t)
\right],
\end{equation}
where $\hat{o}_m(t)$ is the estimated uncovered-user ratio, $\hat{d}_m(t)$ is the estimated RSS deficiency, and $\hat{u}_m(t)$ is the uncertainty of the cell. As shown in Fig.~\ref{framework}, the complete belief map becomes
\begin{equation}
B_t
=
\{b_m(t)\}_{m=1}^{N_g^2}
\in
\mathbb{R}^{N_g\times N_g\times 3}.
\end{equation}

Before any probing, the belief map is initialized with high uncertainty. When a cell is probed, SparsePilot updates the outage and RSS-deficiency channels using the newly measured values:
\begin{equation}
\hat{o}_m(t+1)
=
(1-\eta)\hat{o}_m(t)+\eta\bar{o}_m(t),
\end{equation}
\begin{equation}
\hat{d}_m(t+1)
=
(1-\eta)\hat{d}_m(t)+\eta\bar{d}_m(t),
\end{equation}
where $\eta\in(0,1]$ is the belief update rate. Unprobed cells retain their
previous belief values. In our experiments, we set $\eta=1$, which reduces the update to direct replacement. The uncertainty channel is updated according to the probing count:
\begin{equation}
\hat{u}_m(t)
=
\frac{1}{\sqrt{n_m(t)+1}}.
\end{equation}
Cells that are not probed retain their previous outage and RSS-deficiency estimates and remain uncertain unless they have been probed in previous steps.

In essence, this belief map serves as a compact partial observation of the wireless environment. Unlike the full radio map, it can be constructed from a small number of measurements and explicitly encodes uncertainty, allowing the following control policy to distinguish between regions that are known to be ``weak'' and regions that are simply under-observed.

\begin{algorithm}[t]
\caption{SparsePilot Workflow}
\label{alg:sparsepilot}
\begin{algorithmic}[1]
\STATE \textbf{Input:} probing budget $(K_c,K_u)$, planning horizon $T$, policy $\pi_{\theta}$
\STATE \textbf{Initialize:} cell scores $\{q_m(0)\}$, counts $\{n_m(0)\}$, belief map $B_0$
\FOR{$t=0,\ldots,T-1$}
    \STATE Compute $\mathrm{UCB}_m(t)$ for each occupied cell $m$
    \STATE Select $\mathcal{S}_t$ as the top-$K_c$ occupied cells by $\mathrm{UCB}_m(t)$
    \FOR{each cell $m\in\mathcal{S}_t$}
        \STATE Probe users $\mathcal{P}_m(t)$ with $|\mathcal{P}_m(t)|\le K_u$
        \STATE Compute $\bar{o}_m(t)$, $\bar{d}_m(t)$, and $\tilde{q}_m(t)$
        \STATE Update $q_m(t)$, $n_m(t)$, and $b_m(t)$
    \ENDFOR
    \STATE $s_t \leftarrow [x_t^{\mathrm{UAV}},x^{\mathrm{BS}},\phi_{\mathrm{scene}},t/T,\mathrm{vec}(B_t)]$
    \STATE Sample action $a_t \sim \pi_{\theta}(\cdot|s_t)$
    \STATE Execute $a_t$ in the environment
    \STATE Compute reward $R_t$ using the hidden full wireless state (training/evaluation only)
\ENDFOR
\end{algorithmic}
\end{algorithm}

\subsection{Belief-Guided UAV Mobility Control}
\label{subsec:belief_control}

The sparse-feedback control problem is partially observable because the controller cannot access the full user-level RSS and network-wide state. SparsePilot uses the coverage-based belief map $B_t$ as a compact representation of this hidden state.
We model the resulting belief-guided control problem as a Markov Decision Process (MDP), i.e.
$\mathcal{M}=(\mathcal{S},\mathcal{A},\mathcal{P},R,\gamma)$,
where $\mathcal{S}$ is the belief-guided observation space, $\mathcal{A}$ is the continuous UAV movement action space, $\mathcal{P}$ is the wireless environment transition dynamics induced by UAV mobility, $R$ is the hidden-state reward, and $\gamma$ is the discount factor.

At step $t$, the policy observation is
\begin{equation}
s_t =
\left[
x_t^{\mathrm{UAV}},
x^{\mathrm{BS}},
\phi_{\mathrm{scene}},
t/T,
\mathrm{vec}(B_t)
\right],
\end{equation}
where $x_t^{\mathrm{UAV}}$ denotes the UAV positions, $x^{\mathrm{BS}}$ denotes the base station (BS) positions, $\phi_{\mathrm{scene}}$ denotes scene-level morphology features, $T$ is the planning horizon, and $\mathrm{vec}(B_t)$ is the flattened belief map vector. The scene-level features summarize urban morphology, such as building density, built-up ratio, footprint statistics, and building height statistics.

The action is a continuous UAV displacement command. For $M$ UAVs, the action space can be formulated as
\begin{equation}
a_t =
[\Delta x_1,\Delta y_1,\Delta z_1,\ldots,
\Delta x_M,\Delta y_M,\Delta z_M].
\end{equation}
Each displacement is clipped to the feasible movement range, and the resulting UAV positions are constrained to remain inside the allowed deployment region.

After executing the UAV movement, the training environment computes the hidden RSS and coverage state, which is used only for offline reward computation and evaluation.
Once trained, the policy executes using only sparse wireless measurements and does \textit{not} require online reward computation or access to the dense wireless state.

Let $\Delta C_t=C_t-C_{t-1}$, $\Delta p_t^{k}=p_t^{k}-p_{t-1}^{k}$,
and $\Delta\bar r_t=\bar r_t-\bar r_{t-1}$ denote the stepwise
improvements in coverage, $k$-th percentile RSS, and mean RSS across
users, respectively.
The reward is defined as
\begin{align}
R_t
=&\;
w_c \Delta C_t
+ w_p \Delta p_t^{k}
+ w_{\mu} \Delta \bar{r}_t
+ w_C C_t \nonumber \\
&+
w_P(p_t^{k}-r_{\mathrm{tar}})
- w_L L_t
- w_M M_t
- w_I I_t .
\label{reward}
\end{align}
Here, $p_t^{k}$ is the $k$-th percentile of the users' best RSS values, $\bar{r}_t$ is the mean best RSS across users, $L_t$ is the load variance across serving transmitters, $M_t$ is the normalized UAV movement cost, and $I_t$ is the number of invalid UAV placements. Typically, $k$ is set to 10 in the evaluation study to emphasize tail-user performance.

We use SAC as the continuous-control backend because 3D UAV planning is naturally represented in a continuous action space.
This SAC learns a stochastic policy $\pi_{\theta}(a_t|s_t)$ that maximizes the entropy-regularized return:
\begin{equation}
J(\pi)
=
\sum_{t=0}^{T-1}
\mathbb{E}_{(s_t,a_t)}
\left[
R_t
+
\lambda
\mathcal{H}(\pi_{\theta}(\cdot|s_t))
\right],
\end{equation}
where $\lambda$ is the entropy temperature. During the training stage, each transition contains the belief-guided observation, UAV movement action, hidden-state reward, the next belief-guided observation, and termination indicator. During execution, SparsePilot repeats the same probing-belief-control loop: UCB selects informative probing cells, the probed RSS feedback updates the belief map, and the SAC policy moves the UAVs based on this belief state.
Algorithm~\ref{alg:sparsepilot} summarizes the SparsePilot workflow.

\section{Theoretical Analysis}
\label{sec:analysis}

This section analyzes the sensing-estimation-control loop of SparsePilot. We first characterize the cell-level estimation error of the belief map under sparse wireless measurements. We then show how UCB-guided probing allocates measurements to refine the belief map in coverage-critical or uncertain regions. Finally, we relate belief-map estimation error to the performance of a downstream belief-guided controller.

\subsection{Belief Estimation under Sparse Probing}
\label{subsec:belief_estimation_analysis}

For each of the $N_g^2$ grid cells, SparsePilot maintains a probing count $n_m(t)$ and an empirical deficiency estimate $q_m(t)$ based on the bounded instantaneous score $\tilde{q}_m(t)\in[0,1]$ defined in Eq.~(\ref{badness}). This estimate essentially summarizes the coverage-deficiency information encoded in the RSS-deficiency channels of the belief map. Let $\mu_m=\mathbb{E}[\tilde{q}_m(t)]$ denote the expected local deficiency of cell $m$. Because UAV movements alter the wireless state, we assume local stationarity over a short probing-control interval, within which $\mu_m$ is treated as approximately fixed.

\textbf{Lemma 1.} \emph{For any confidence parameter $\delta\in(0,1)$, with probability at least $1-\delta$, every grid cell with $n_m(t)\geq 1$ satisfies}
\begin{equation}
|q_m(t)-\mu_m|
\le
\sqrt{
\frac{
\log(2N_g^2/\delta)
}{
2n_m(t)
}}.
\end{equation}

\textit{Proof.}
For a fixed cell $m$ with $n_m(t)\geq 1$, $q_m(t)$ is the average of $n_m(t)$ bounded observations in $[0,1]$. By Hoeffding's inequality \cite{hoeffding1963probability},
\begin{equation}
\Pr\left(|q_m(t)-\mu_m|\ge \epsilon\right)
\le
2\exp(-2n_m(t)\epsilon^2).
\end{equation}
Setting the right-hand side to $\delta/N_g^2$ and applying a union bound over all probed cells gives the result.
\hfill $\square$

Lemma 1 shows that the cell-level belief error decreases as more wireless measurements are allocated to that cell. Therefore, the quality of the complete belief map depends not only on the total probing budget, but also on how measurements are distributed across the spatial regions.

\subsection{UCB-Guided Belief Map Refinement}
\label{subsec:ucb_analysis}

SparsePilot uses the UCB rule in Eq.~(\ref{ucb}) to allocate the sparse probing budget across grid cells. 
Let $\mathcal{S}^{\star}$ denote the set of $K_c$ cells with the largest expected local deficiency:
\begin{equation}
\mathcal{S}^{\star}
=
\arg\max_{\mathcal{S}:|\mathcal{S}|=K_c}
\sum_{m\in\mathcal{S}}\mu_m.
\end{equation}
Let $\mathcal{S}_t$ be the set of cells selected at step $t$. We define a local measurement allocation regret as
\begin{equation}
R_T^{\mathrm{probe}}
=
\sum_{t=1}^{T}
\left(
\sum_{m\in\mathcal{S}^{\star}}\mu_m
-
\sum_{m\in\mathcal{S}_t}\mu_m
\right).
\end{equation}
This regret measures the sensing inefficiency of the selected probing cells relative to the oracle top-$K_c$ cells with the largest expected local deficiency scores. It is not a control regret for the learned UAV mobility policy.

\textbf{Lemma 2.} \emph{Under the local stationarity view above, the UCB probing rule satisfies}
\begin{equation}
\mathbb{E}
\left[
R_T^{\mathrm{probe}}
\right]
=
O
\left(
\sum_{m\notin\mathcal{S}^{\star}}
\frac{\log T}{\Delta_m}
\right),
\end{equation}
\emph{where $\Delta_m=\mu_{(K_c)}-\mu_m>0$ is the gap between cell $m$ and the $K_c$-th largest local deficiency score.}

\textit{Proof sketch.} The probing problem is a multiple-play bandit over occupied grid cells, where each selected cell returns a bounded deficiency observation. Standard UCB analysis implies that each suboptimal cell is selected only logarithmically many times in expectation \cite{auer2002finite}. Extending this argument to top-$K_c$ selection gives the stated bound. \hfill $\square$

Lemma 2 shows that UCB balances exploitation of high-deficiency cells with exploration of insufficiently observed cells. In contrast, random probing allocates measurements indiscriminately, while greedy probing may repeatedly sample already well-characterized weak regions and neglect uncertain cells.
Thus, our UCB policy improves the belief map selectively, as it allocates the limited measurement budget to regions that are either likely to contain coverage deficiencies or remain poorly characterized.

\subsection{Impact of Belief Map Error on Downstream Control}
\label{subsec:control_gap_analysis}

We next relate cell-level deficiency estimation error to the value of a
belief-guided SAC-based policy. Let $B$ denote the ideal cell-level deficiency
representation constructed from the expected local deficiencies
$\{\mu_m\}$, and let $\hat{B}$ denote its estimate obtained from sparse
probing. The policy acts on $\hat{B}$, while the hidden full wireless state
is used only during offline training and evaluation. We adopt the following smoothness assumption:

\textbf{Assumption 1.}
\emph{For a fixed learned policy $\pi$, there exists a local region of
belief representations encountered during execution such that, for any
$B$ and $\hat{B}$ in this region,}
\begin{equation}
\left|V^{\pi}(B)-V^{\pi}(\hat{B})\right|
\leq
L_B
\|B-\hat{B}\|_{1,w},
\label{belief_smoothness}
\end{equation}
\emph{where $L_B$ is a belief-to-value smoothness constant and}
\begin{equation}
\|B-\hat{B}\|_{1,w}
=
\sum_{m=1}^{N_g^2}
w_m |B_m-\hat{B}_m|,
\end{equation}
\emph{with $w_m\geq 0$ and $\sum_{m=1}^{N_g^2}w_m=1$.}

This assumption does not require the learned policy to be globally optimal; it only imposes a smoothness bound on how its value changes with the belief representation.

\textbf{Theorem 1.} \emph{Under Assumption 1, with probability at least $1-\delta$, the performance gap between executing the same control policy $\pi$ using the ideal belief $B$ and using the estimated belief $\hat B$ satisfies}
\begin{equation}
\left|V^{\pi}(B)-V^{\pi}(\hat{B})\right|
\le
L_B
\sum_{m=1}^{N_g^2}
w_m
\sqrt{
\frac{
\log(2N_g^2/\delta)
}{
2n_m(t)
}},
\end{equation}
\emph{where $w_m\geq 0$ is the importance weight of cell $m$,
$\sum_{m=1}^{N_g^2}w_m=1$, and $n_m(t)\geq 1$. The weight $w_m$ can
reflect user density, outage severity, or the contribution of cell $m$
to the policy observation.}

\textit{Proof.}
Applying Lemma 1 to each cell and using Assumption 1 in
Eq.~(\ref{belief_smoothness}), we obtain
\begin{equation}
\|B-\hat{B}\|_{1,w}
\leq
\sum_{m=1}^{N_g^2}
w_m
\sqrt{
\frac{\log(2N_g^2/\delta)}
{2n_m(t)}
}.
\end{equation}
Substituting this bound into the local smoothness inequality completes
the proof.
\hfill $\square$

Theorem 1 connects sparse sensing to the quality of the belief map and, consequently, to downstream belief-guided control. The bound decreases as high-importance cells receive more probing samples, meaning that active probing improves the belief representation in regions that matter most to the network planning objective. This result is not tied to a particular DRL framework: any downstream controller whose value is locally smooth with respect to the belief map can benefit from a more accurate and task-relevant belief representation.

\section{Evaluation}

\subsection{Experimental Setup}
\label{subsec:exp_setup}

\textbf{Scenario configuration.}
We evaluate SparsePilot in seven real-map-based urban wireless DTs, including Raleigh, Atlanta, Chicago, Seattle, Boston, San Francisco, and Los Angeles. The urban geometries are derived from OpenStreetMap data~\cite{openstreetmap}, converted into ray-tracing-compatible scene meshes, and loaded into Sionna RT~\cite{sionna} for evaluation. For each episode, the DT places ground users, terrestrial base stations, and UAV-mounted transmitters in the target scene and computes user-level RSS values from Sionna-generated radio maps. We use the same RF and ray-tracing configuration across all scenes unless otherwise stated. Table~\ref{tab:wireless_config} summarizes the main wireless and ray-tracing parameters. The policy is trained only in Raleigh and evaluated on Raleigh and the six unseen urban scenarios.

\begin{table}[t]
\centering
\caption{RF and Sionna ray-tracing configuration.}
\label{tab:wireless_config}
\begin{tabular}{l l}
\toprule
\textbf{Parameter} & \textbf{Value} \\
\midrule
Maximum ray depth & 2 \\
Ray-tracing samples & 100{,}000 per transmitter \\
Radio-map cell size & $5\,\mathrm{m} \times 5\,\mathrm{m}$ \\
Ground-user height & $1.5\,\mathrm{m}$ \\

Carrier frequency & 3.5 GHz \\
Antenna model & TR 38.901, vertical polarization \\

BS \& UAV transmit power & $44\,\mathrm{dBm}$ \\
Coverage threshold $\tau$ & $-90\,\mathrm{dBm}$ \\
Target RSS $r_{\mathrm{tar}}$ & $-80$ dBm \\

Propagation effects & LoS, reflection, refraction \\
RSS clipping range & $[-140,-20]$ dBm \\
Coverage metric & Strongest RSS over UAVs and BSs \\
\bottomrule
\end{tabular}
\vspace{-1.5em}
\end{table}

\textbf{Mobility configuration.}
Each episode contains $N_u=150$ ground users (GUs), scene-specific BSs, and $M=5$ UAV-mounted radios, and lasts for $T=5$ planning steps. Most cities use two BSs, while Boston uses one BS.
At each step, the 3D displacement of each UAV is clipped to $\Delta x,\Delta y\in[-250,250]$m and $\Delta z\in[-40,40]$m. The resulting UAV positions are constrained to the deployment region $x,y\in[-1000,1000]$m and $z\in[50,220]$m. A placement is invalid if it lies within a building footprint and below the corresponding building height plus a $1$m safety margin.

\textbf{Training configuration.}
For sparse feedback, we use an $N_g\times N_g$ grid with $N_g=10$ and probe at most $K_c=3$ cells with up to $K_u=2$ users per selected cell at each planning step.
We set the UCB exploration coefficient to $\beta=1.0$. The normalized RSS-deficiency term uses $D_{\max}=60$~dB, the cell badness score uses $\alpha=0.7$, and the belief map
update rate is set to $\eta=1$.
Before probing, the outage, RSS-deficiency, and uncertainty channels are initialized to $0.5$, $0$, and $1$, respectively.

We train the feedback-driven SAC policies in Raleigh for $20{,}000$
environment steps. Both the actor and critic use two fully connected hidden
layers with $256$ units per layer and ReLU activations. We use Adam with a
learning rate of $3\times10^{-4}$, a batch size of $256$, a replay buffer of
$10^6$ transitions, discount factor $\gamma=0.95$, soft-update coefficient
$\tau_{\mathrm{SAC}}=0.005$, and automatic entropy tuning.
For the reward function in Eq.~(\ref{reward}), we set $w_c=20$,
$w_p=0.20$, $w_{\mu}=0.05$, $w_C=2.0$, $w_P=0.03$, $w_L=0.02$,
$w_M=0.01$, and $w_I=30$. The same hidden-state reward is used for all learning-based methods to ensure a fair comparison.

\textbf{Baselines.}
We compare SparsePilot with the following baselines:
\begin{itemize}
\item \textbf{Static}: All UAVs remain at their initial deployment locations.
\item \textbf{Random}: The UAV randomly samples feasible movement actions at each planning step.
\item \textbf{Greedy}: A full-information heuristic that identifies currently uncovered GUs at each step and moves all UAVs toward their horizontal centroid under the same movement constraints, with altitudes unchanged.
\item \textbf{Raw SAC}: A standard SAC baseline that uses complete user RSS and coverage observations, without the proposed belief-map representation.
\item \textbf{SparsePilot w/o UCB}: A sparse-feedback SAC controller that uses random probing.
\item \textbf{SparsePilot w/ $\epsilon$-Greedy}: A SparsePilot variant that replaces UCB probing with $\epsilon$-greedy probing while retaining the same belief state and SAC controller.
\end{itemize}

\begin{figure}[t]
  \centering
  \includegraphics[width=0.312\linewidth]{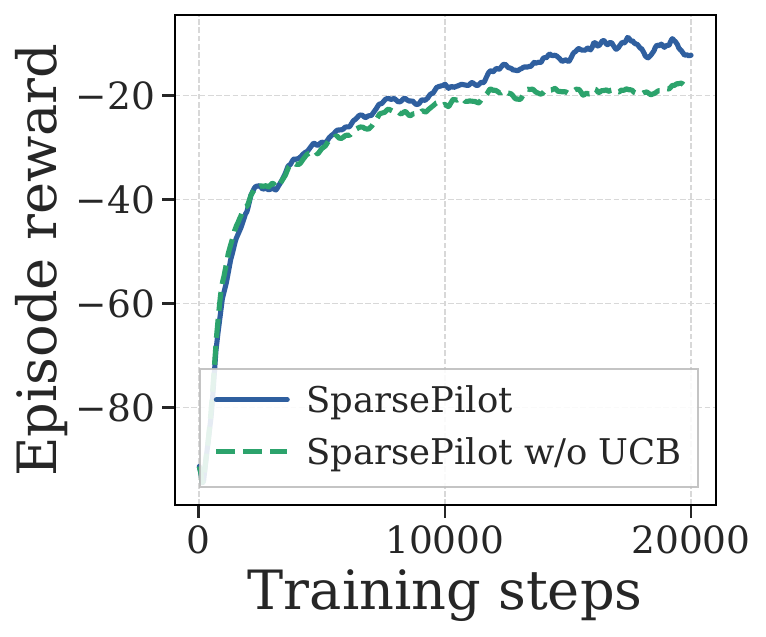}
  \hspace{-0.7em}
  \includegraphics[width=0.312\linewidth]{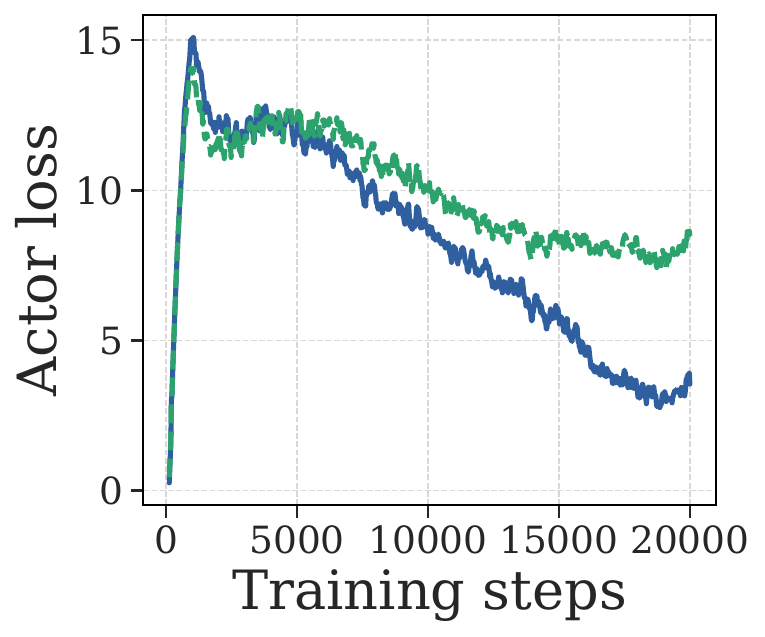}
  \hspace{-0.7em}
  \includegraphics[width=0.312\linewidth]{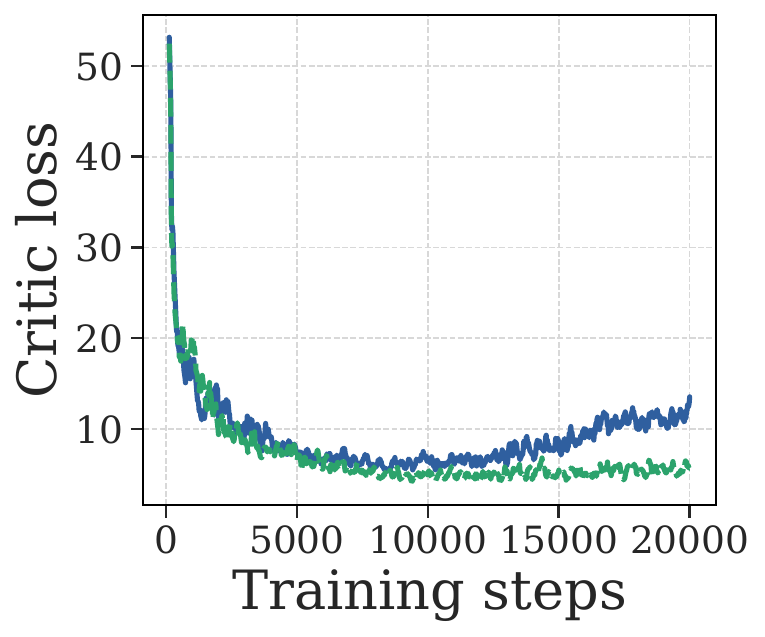}
    \vspace{-0.5em}
  \caption{\small Training dynamics of SparsePilot with UCB-guided probing and random probing.}
  \label{training}
  \vspace{-1.5em}
\end{figure}

\textbf{Evaluation metrics.}
Each method is evaluated over $20$ random episodes per city. We report the mean and standard deviation across episodes. The primary metric is the network coverage ratio, defined as the fraction of covered GUs after the planning. We also report coverage gain, mean RSS across users, episode reward, load fairness, and the measurement cost.

\textbf{Training behavior.}
Fig.~\ref{training} shows the training curves of SparsePilot with UCB probing and random probing in the Raleigh scenario. Both policies become stable within $20{,}000$ environment steps, indicating that the belief-guided observation provides a learnable state representation for continuous planning. Compared with random probing, UCB probing achieves a higher final episode reward, suggesting that actively selected measurements provide more informative belief states for policy learning. The actor and critic losses remain bounded after the early training stage, showing a stable learning process under sparse feedback.

\subsection{Cross-City Deployment Performance}

\begin{figure}[t]
  \centering
  \vspace{-0.3cm}
  \includegraphics[width=0.9\linewidth]{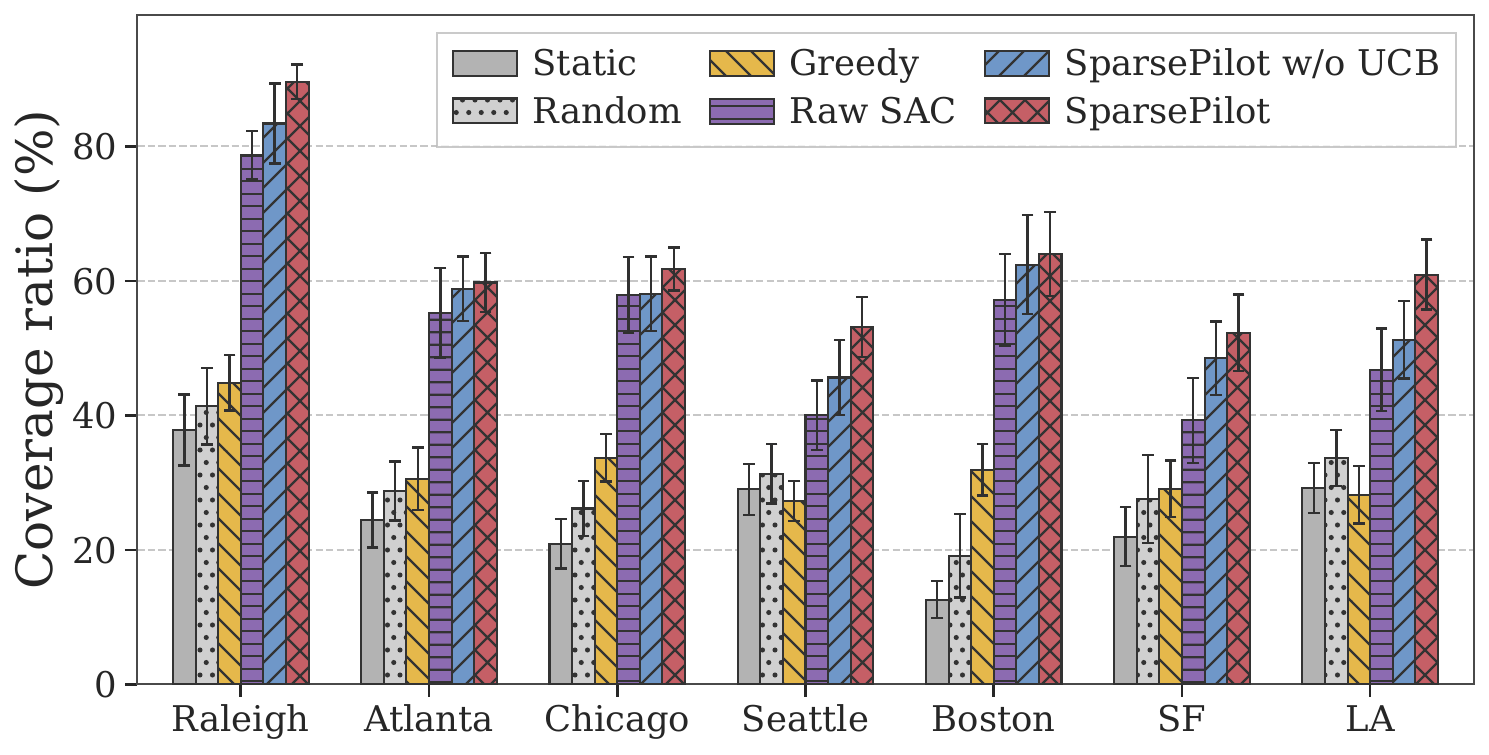}
  \vspace{-0.6em}
  \includegraphics[width=0.9\linewidth]{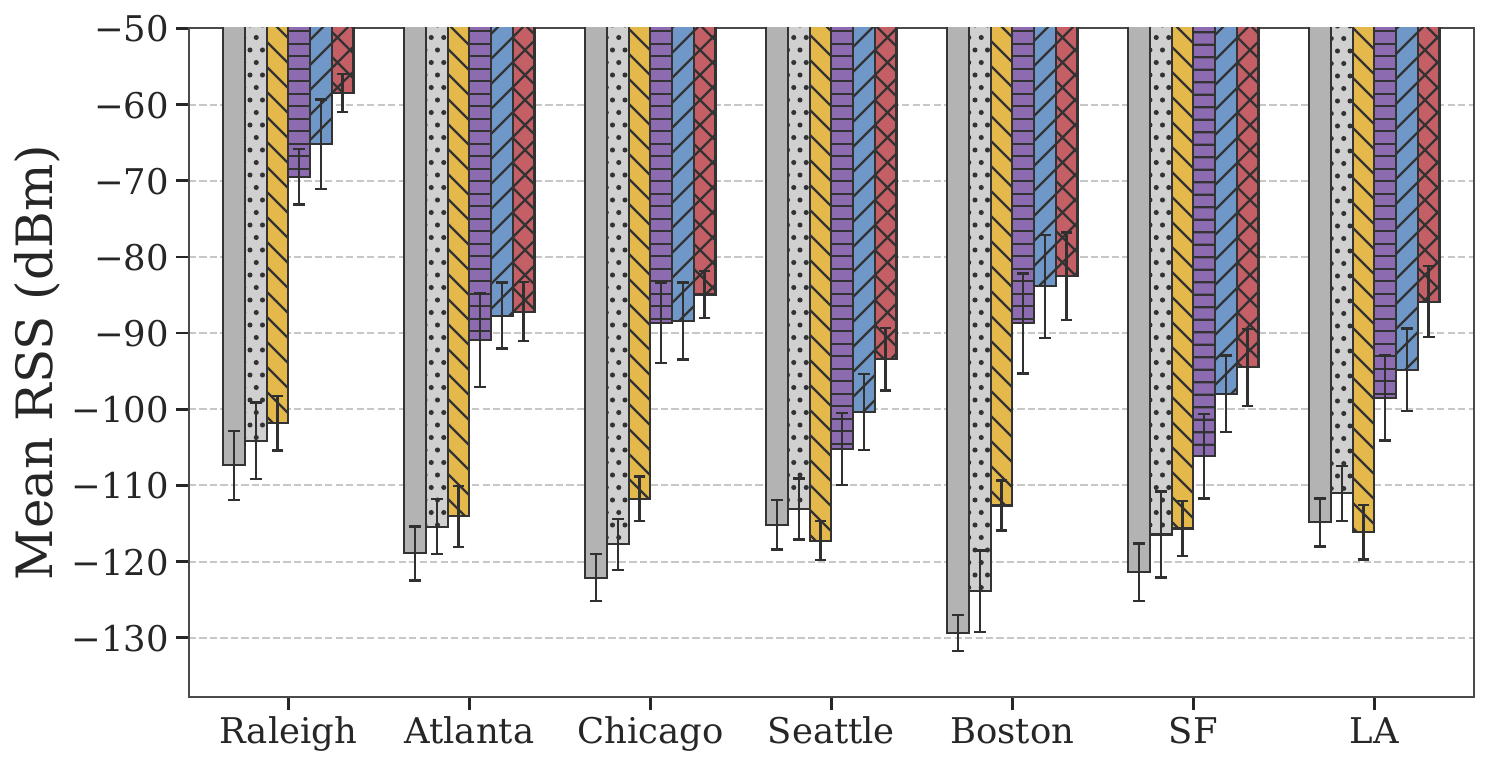}
  \caption{\small Cross-city coverage restoration performance.}
  \label{coverage}
  \vspace{-2em}
\end{figure}

We first evaluate whether the Raleigh-trained planning policy can generalize to unseen cities with different urban morphologies under sparse wireless feedback.
Fig.~\ref{coverage} reports the coverage ratio and mean RSS across all seven cities. SparsePilot achieves the highest average coverage ratio of $63.08\%$, compared with $58.30\%$ for SparsePilot without UCB, $53.59\%$ for Raw SAC, and $32.23\%$ for the greedy heuristic. It also provides the largest average coverage gain and mean RSS across users, improving coverage by $37.93$ percentage points from the initial deployment. The consistent improvements across both the source and unseen target cities indicate that the learned SAC-based controller transfers effectively across different urban morphologies and propagation conditions.

More importantly, SparsePilot uses only $23.21$ user measurements per episode on average. Since full observation would require $150\times5=750$ user-step RSS measurements, SparsePilot uses only about $3.1\%$ of the full-observation measurement budget. 
The comparison with Raw SAC further highlights the role of state representation. Raw SAC is a full-observation baseline with the same SAC backbone, but it is not intended to represent the strongest possible full-state architecture. It directly consumes dense GU-level observations that are high-dimensional and dependent on the sampled user layout, whereas SparsePilot abstracts sparse measurements into a fixed-size, spatially aligned belief map encoding outage, RSS deficiency, and uncertainty.

\textbf{Remark.} SparsePilot's cross-city generalization benefits from its active sensing and task-oriented belief representation. During deployment, UCB acquires informative measurements from the current environment, and the belief map summarizes coverage deficiency and uncertainty in a fixed-size spatial representation. This allows the SAC policy to make decisions based on updated belief states rather than scene-specific dense user observations, helping SparsePilot transfer to unseen urban environments without online retraining. This may help explain the superior transfer performance of SparsePilot relative to Raw SAC across unseen cities.

\subsection{Impact of Sparse-Feedback Budget}

\begin{figure}[t]
  \centering
  \includegraphics[width=0.45\linewidth]{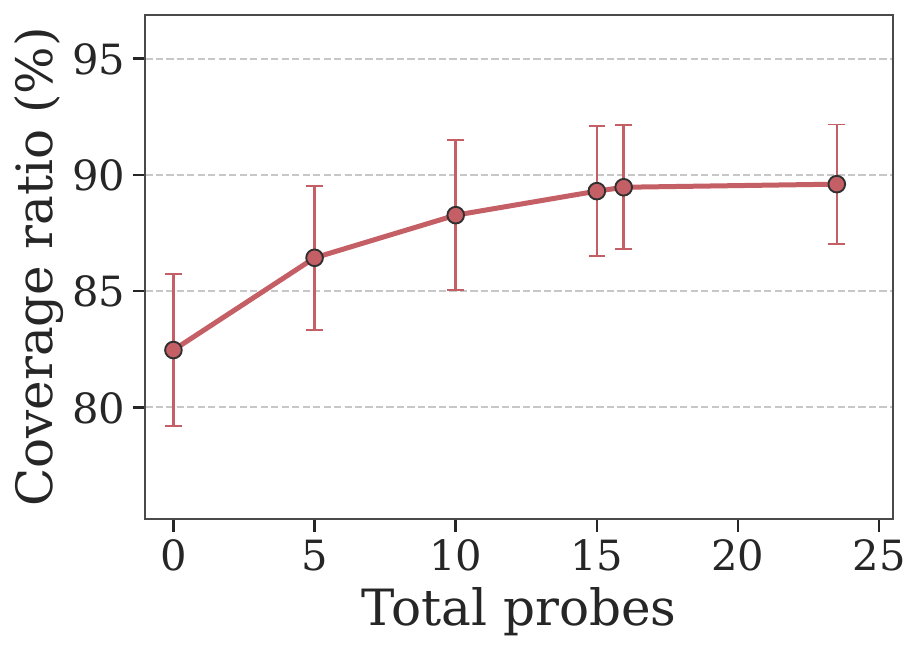}
  \hspace{-0.3em}
  \includegraphics[width=0.45\linewidth]{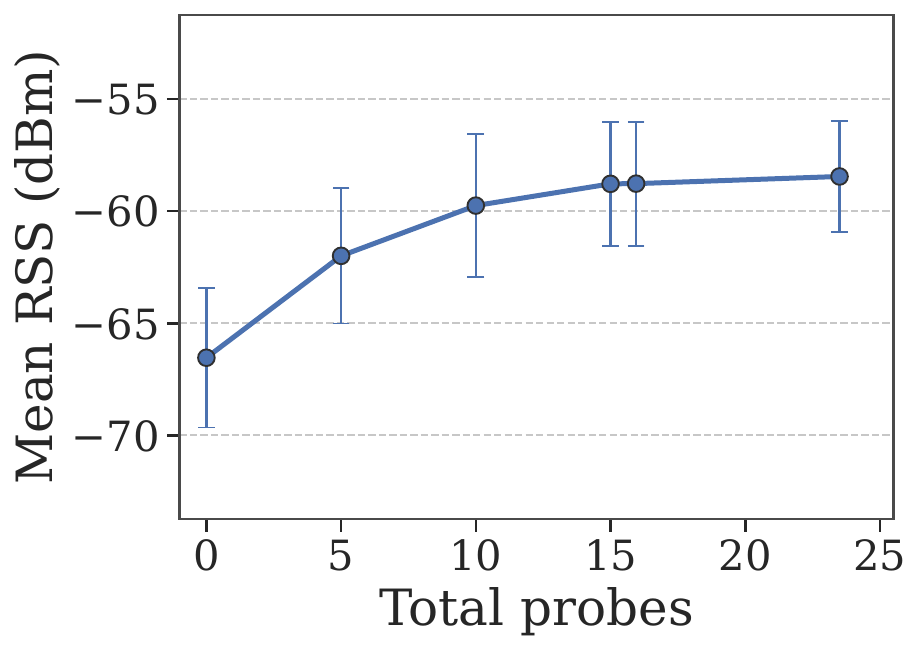}
    \vspace{-0.5em}
  \caption{\small Impact of the probing budget in Raleigh.
  }
  \label{Budget}
  \vspace{-1.5em}
\end{figure}

We study the tradeoff between wireless measurement cost and coverage improvement by varying the per-step probing budget under $T=5$ and varying the planning horizon under a fixed budget of three cells and two users per cell. Fig.~\ref{Budget} shows that even a small number of targeted
measurements substantially improves coverage and mean RSS in Raleigh. Performance increases rapidly with the first few probes and then exhibits diminishing returns after approximately 16 probed users. With around 23.5 probes per episode, SparsePilot reaches an $89.60\%$ coverage ratio, indicating that most of the feedback benefit can be obtained with only a few tens of measurements.

\begin{figure}[t]
  \centering
    \vspace{-0.3cm}
  \subfloat[Average.\label{fig:deadline}]{
    \includegraphics[width=0.315\linewidth]{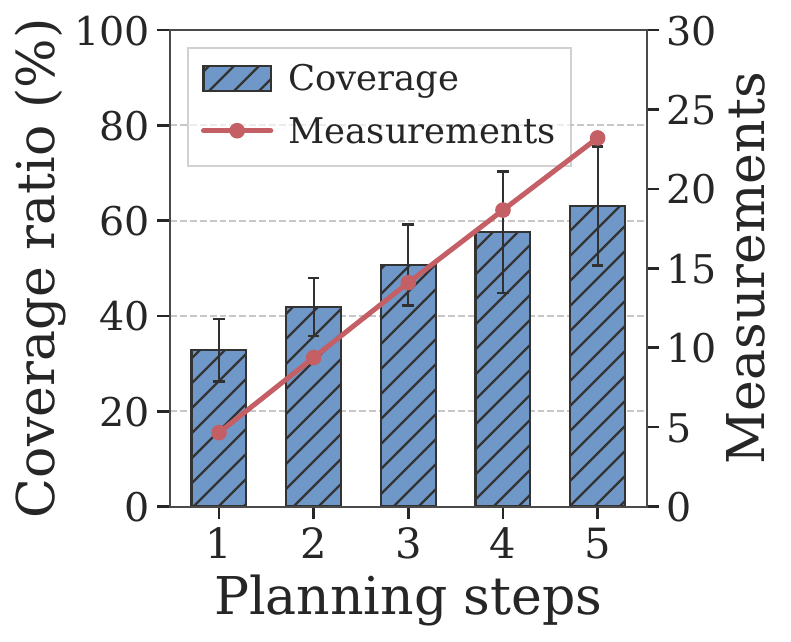}
  }
      \hspace{-0.7em}
  \subfloat[Raleigh.\label{fig:avg_e2e}]{
    \includegraphics[width=0.315\linewidth]{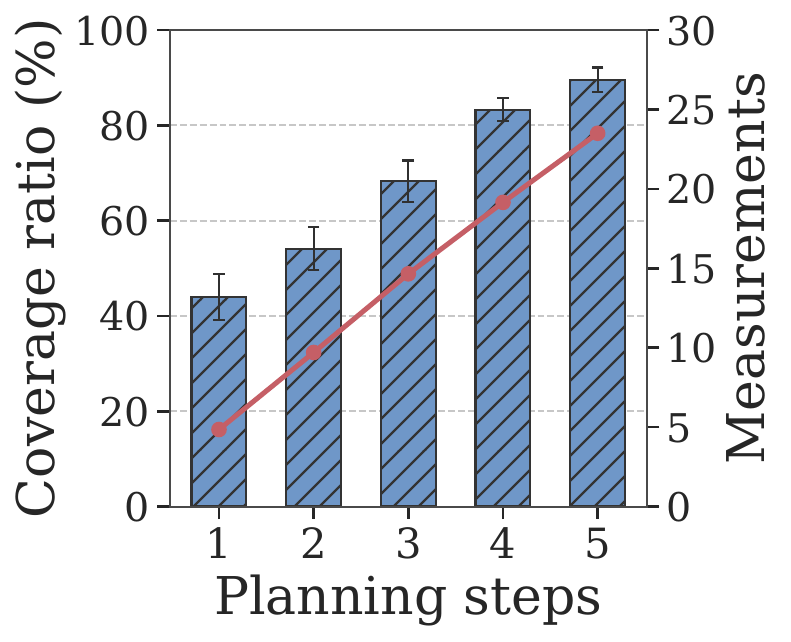}
  }
  \hspace{-0.7em}
  \subfloat[Boston.\label{fig:goodput}]{
    \includegraphics[width=0.315\linewidth]{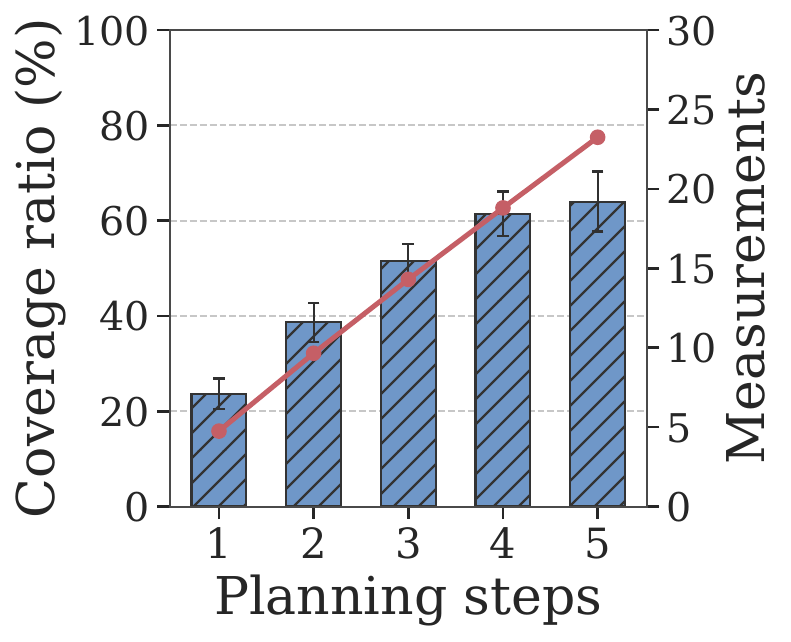}
  }
    \vspace{-0.5em}
  \caption{\small Impact of the planning horizon on final coverage and measurement count. }
  \label{steps}
  \vspace{-0.35em}
\end{figure}

Fig.~\ref{steps} further shows that additional planning steps improve coverage through iterative sensing-control refinement. As the planning horizon increases from $T=1$ to $T=5$, the cross-city average coverage ratio increases from $32.81\%$ to $63.08\%$, while the average measurement count increases from $4.65$ to $23.21$ per episode. Each additional round provides more opportunities to update the belief map and progressively mitigate weak-coverage regions. This behavior follows the implication of Lemma~1 in our theoretical analysis: additional sparse measurements reduce the estimation error of probed belief-map cells, providing a more reliable state representation for downstream control decisions.

\subsection{Robustness under Imperfect Feedback}
\label{subsec:robustness}

\begin{figure}[t]
  \centering
\hspace{-0.8em}
  \includegraphics[width=0.33\linewidth]{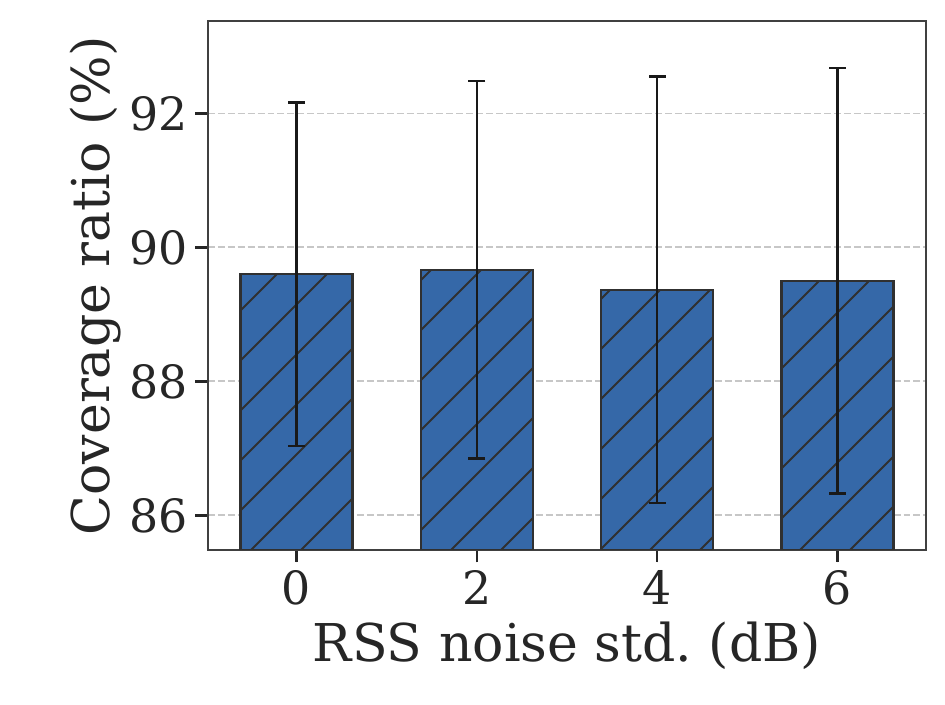}
  \hspace{-0.5em}
  \includegraphics[width=0.33\linewidth]{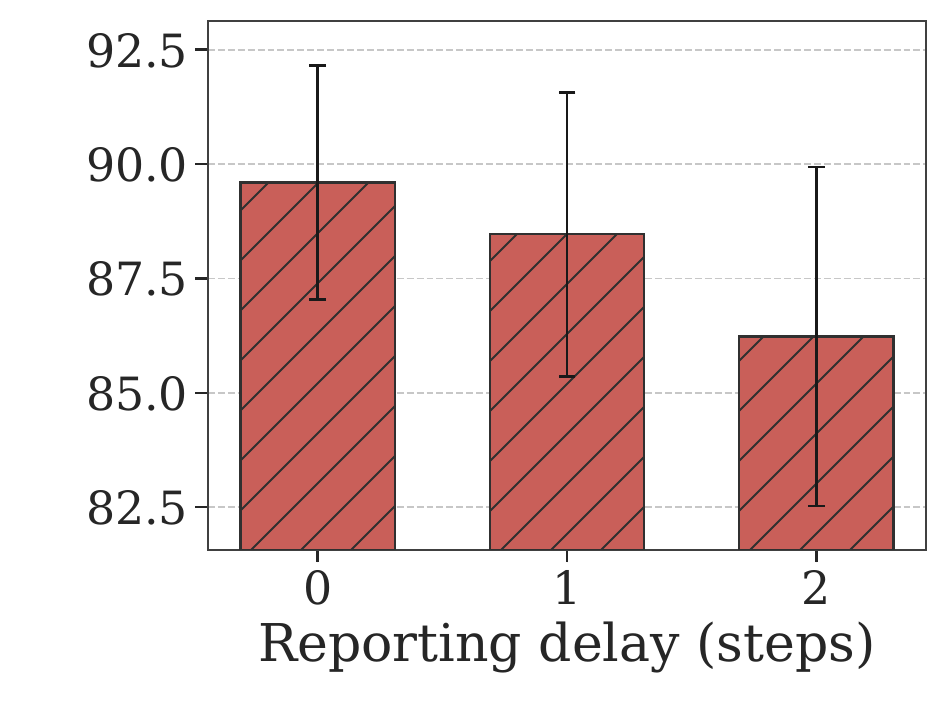}
  \hspace{-0.7em}
  \includegraphics[width=0.33\linewidth]{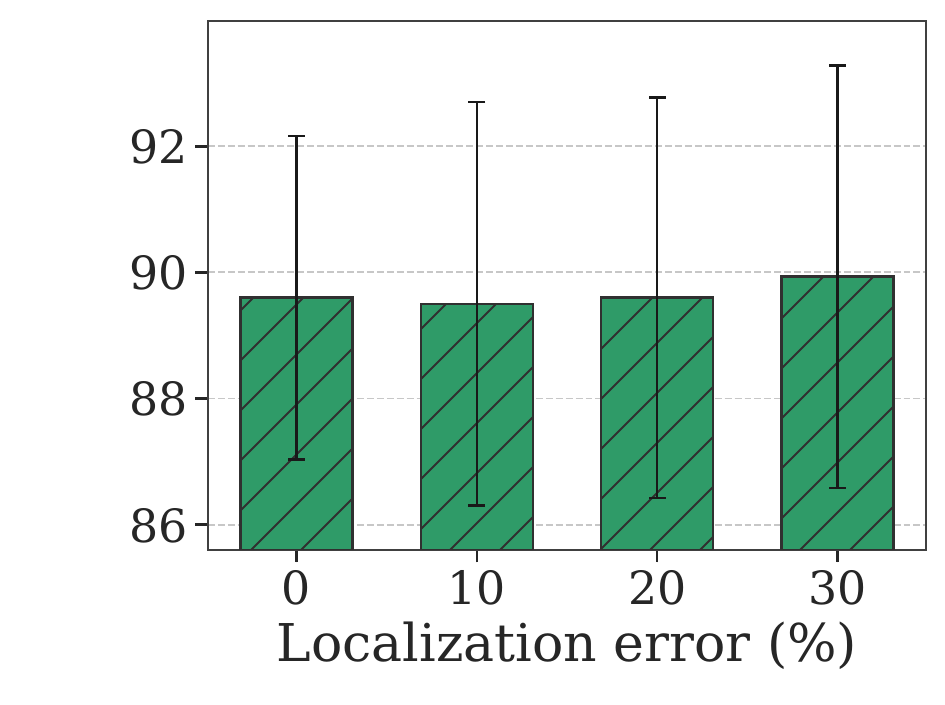}

  \vspace{-0.5em}
  \caption{\small Robustness to practical feedback imperfections.}
  \label{fig:robustness}
\end{figure}

We further evaluate SparsePilot in Raleigh under three practical feedback imperfections: noisy RSS measurements, reporting delay, and coarse user-localization errors. Each setting uses 20 random episodes and varies one imperfection at a time while keeping the other two at the clean setting. As shown in Fig.~\ref{fig:robustness}, SparsePilot is robust to RSS measurement noise. Increasing the RSS noise standard deviation from $0$ to $6$ dB changes the final coverage only from $89.60\%$ to $89.50\%$, with the largest drop being $0.23$ percentage points at $4$ dB noise. SparsePilot is also insensitive to neighboring-cell localization errors: with up to $30\%$ localization errors, the coverage ratio remains between $89.50\%$ and $89.93\%$, and the small non-monotonic variations are within episode-level randomness. Reporting delay has the largest impact because delayed channel measurements make the belief map less synchronized with the current UAV deployment. A two-step delay reduces the coverage ratio from $89.60\%$ to $86.23\%$ and decreases mean RSS by $3.90$ dB. Nevertheless, SparsePilot still maintains high coverage under delayed feedback, indicating that our sensing-estimation-control loop remains effective under moderate feedback imperfections.

\subsection{Effectiveness of UCB Active Probing}

To investigate the impact of active probing, we compare SparsePilot with two sparse-feedback variants under the same probing budget. The first variant, SparsePilot w/o UCB, uses random probing during both training and evaluation. The second variant replaces UCB with a $\epsilon$-greedy probing rule at test time, while keeping the learned movement controller unchanged.
Table~\ref{tab:bandit_ablation} shows the average performance over seven cities. Random probing achieves an average coverage ratio of $58.30\%$, while $\epsilon$-greedy probing improves it to $61.49\%$. UCB probing further increases the performance to $63.08\%$, achieving a $4.77$ percentage-point gain over random probing and a $1.59$ percentage-point gain over $\epsilon$-greedy probing. The same trend is observed in coverage gain.

The improvement is attributed to the exploration--exploitation balance of UCB. Random probing may allocate measurements to well-served or uninformative regions, while $\epsilon$-greedy probing may repeatedly exploit known weak regions and overlook uncertain areas. 
These results are consistent with Lemma~2 in Sec.~\ref{subsec:ucb_analysis}: by combining estimated coverage deficiency with an uncertainty bonus, UCB refines the belief map in regions that are likely to be under-covered or insufficiently explored. The improved coverage and RSS further support the implication of Theorem~1 that a more accurate and task-relevant belief map can benefit the downstream controller under the same measurement budget.

\begin{table}[t]
\centering
\caption{Effect of probing strategies averaged across seven cities.
Values report the mean and standard deviation across city-level
averages.}
\label{tab:bandit_ablation}
\begin{tabular}{lcc}
\toprule
Probing Strategy
& Coverage Ratio (\%)
& Mean RSS (dBm) \\
\midrule
Random
& $58.30 \pm 12.61$
& $-88.33 \pm 11.79$ \\
$\epsilon$-Greedy
& $61.49 \pm 13.51$
& $-85.29 \pm 12.89$ \\
UCB
& $\mathbf{63.08 \pm 12.49}$
& $\mathbf{-83.84 \pm 12.03}$ \\
\bottomrule
\end{tabular}
\vspace{-1em}
\end{table}

\subsection{Load Fairness and Service Balance}

Besides improving coverage, a practical planning policy should also maintain balanced service across radio transmitters. We evaluate load balancing using Jain's fairness index \cite{jain1984quantitative} and load variance. Let $\ell_j$ denote the number of GUs associated with serving transmitter $j$, and let $M_s$ be the number of serving transmitters. Jain's fairness index is defined as $J_{\mathrm{fair}}=(\sum_{j=1}^{M_s}\ell_j)^2/ (M_s\sum_{j=1}^{M_s}\ell_j^2)$.

\begin{figure}[t]
  \centering
  \includegraphics[width=0.47\linewidth]{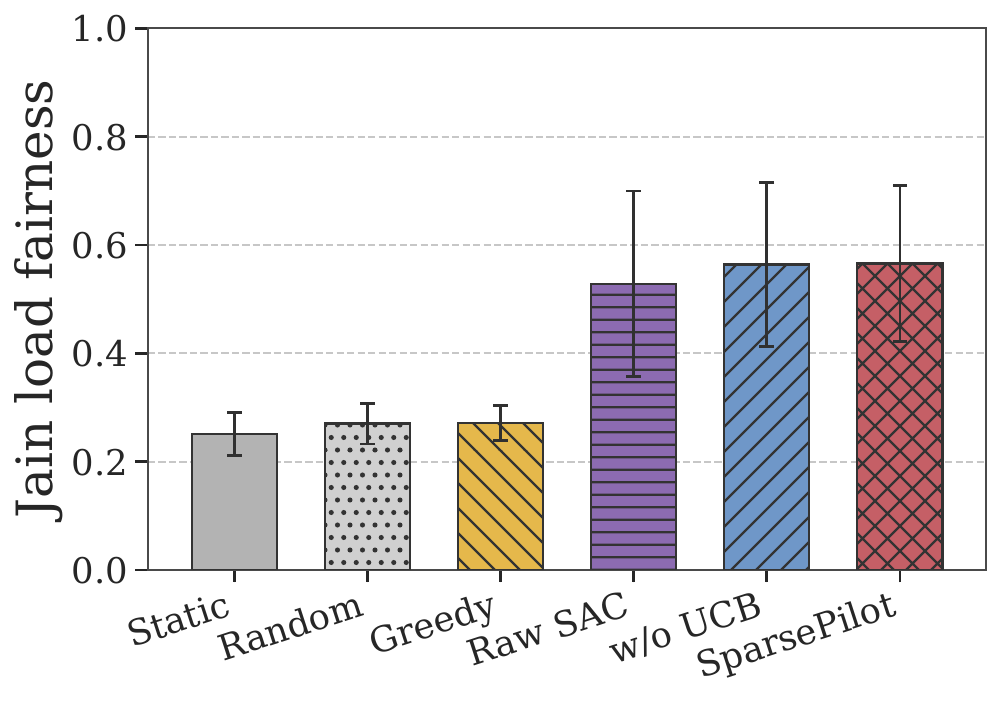}
  \hspace{-0.2em}
  \includegraphics[width=0.47\linewidth]{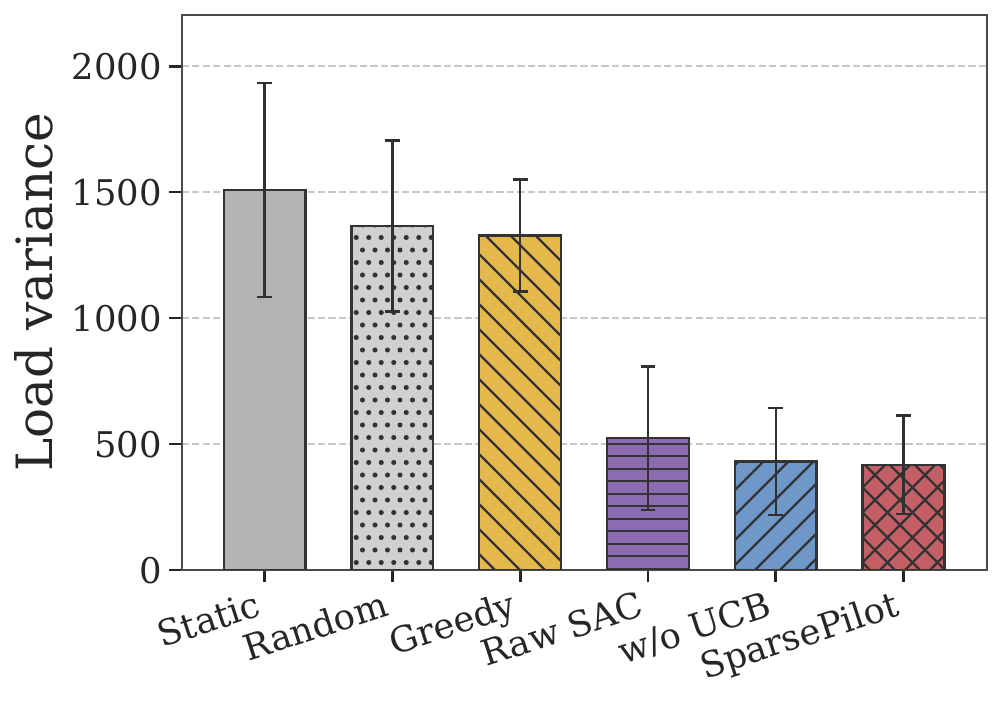}
    \vspace{-0.5em}
  \caption{\small Load fairness comparison across cities.}
  \label{Fairness}
  \vspace{-1.5em}
\end{figure}

Fig.~\ref{Fairness} shows the cross-city average performance. SparsePilot achieves the highest Jain fairness and the lowest load variance among all baselines, indicating that our planning method can improve both coverage and service balance. SparsePilot and SparsePilot without UCB obtain similar fairness, suggesting that load distribution is mainly determined by the learned control policy, while UCB-guided probing improves coverage \textit{without} degrading service balance.

\section{Related Work}
\label{sec:related}

\textbf{Radio Map Construction and Sparse Wireless Sensing.}
Radio maps and channel knowledge maps have been widely studied as environment-aware representations for wireless network planning, localization, spectrum management, and resource optimization \cite{feng2025recent}. Existing radio map estimation methods can be broadly categorized into model-driven, data-driven, and hybrid approaches \cite{bi2019engineering}, with recent learning-based methods using CNNs \cite{levie2021radiounet}, Transformers \cite{viet2025spatial}, and graph neural networks \cite{chen2023graph} to infer dense RSS or channel maps from environmental layouts and sparse measurements. 
More recent methods further improve sparse radio map reconstruction by using generative models \cite{luo2025denoising, liu2020cgan}, multi-scale feature fusion, or physics-aware priors~\cite{shahid2025reveal}. Active sensing has also been explored for radio map construction, where uncertainty-aware UAV trajectories are planned to collect informative measurements and accelerate dense-map reconstruction~\cite{shrestha2022spectrum}.
Unlike prior work that aims to reconstruct full radio maps, SparsePilot builds a \textit{compact belief map} from sparse measurements for cost-effective network control.

\textbf{UAV Planning under Partial Observability.}
Planning under partial observability is commonly formulated as a partially observable Markov decision process, in which the controller cannot directly access the full system state and must act from incomplete observations \cite{kaelbling1998planning, spaan2012partially}. A common approach is to maintain a belief representation that aggregates prior knowledge and sequential observations into a compact decision state \cite{karkus2017qmdp, zhou2026beliefmapnav}. Alternatively, recurrent policies can implicitly encode observation histories when an explicit belief state is difficult to construct \cite{hausknecht2015deep, luo2024efficient}.
Recent studies have applied these ideas to UAV network planning. For example, \cite{ye2022multi} combines graph attention and recurrent memory to aggregate neighboring and historical observations, while \cite{gao2025csmaac} selectively communicates with influential UAVs and incorporates a safety layer for collision avoidance. However, these methods assume a \textit{fixed} local observation process and primarily mitigate incomplete information through recurrent memory or inter-UAV communication, rather than selectively performing wireless measurements under a limited feedback budget. In contrast, SparsePilot couples active wireless probing with UAV mobility control by acquiring sparse RSS measurements to maintain a wireless coverage belief that directly guides continuous network planning.

\section{Conclusion}
\label{sec:conclusion}

This paper studied UAV-assisted network planning under sparse wireless feedback. SparsePilot integrates UCB-guided active probing,  coverage belief construction, and belief-guided continuous control, enabling closed-loop adaptation without exposing the full wireless state to the policy. Our theoretical analysis connects probing frequency, belief estimation error, and the sparse-feedback control gap. Experiments across seven urban digital twins showed that SparsePilot achieves the highest average coverage restoration while
using only about 3.1\% of the full-observation measurement budget. These results also demonstrate strong cross-scene generalization to unseen urban-scale wireless network environments.

\newpage

\bibliographystyle{IEEEtran}
\bibliography{reference}
\end{document}